\documentclass[12pt]{article}
\usepackage{amsfonts}
\newcommand{\ret}{\nonumber \\}
\newcommand{\Section}[1]%
{\section{#1}\setcounter{equation}{0}%
\setcounter{theorem}{0}}

\def\av{\mathbb{E}}
\def\ze{\mathbb{Z}}

\def\sp{\mathbb{S}}
{\par\noindent{\em #1:\ }}%
{~\rule{2mm}{2mm}\par\bigskip}
\begin{document}
\newpage
\thispagestyle{empty}
{\topskip 2cm
\begin{center}
{\Large\bf Fluctuations Destroying Long-Range Order\\}
\bigskip
{\Large\bf in SU(2) Yang-Mills Theory\\}
\bigskip\bigskip\bigskip
{\large Tohru Koma}\\
\bigskip
{\small \it Department of Physics, Gakushuin University, 
Mejiro, Toshima-ku, Tokyo 171-8588, JAPAN}\\
\smallskip
{\small\tt e-mail: tohru.koma@gakushuin.ac.jp}
\end{center}
\vfil
\noindent
We study lattice SU(2) Yang-Mills theory with dimension $d\ge 4$. The model 
can be expressed as a $(d-1)$-dimensional O(4) non-linear $\sigma$-model 
in a $d$-dimensional heat bath. As is well known, the non-linear 
$\sigma$-model alone shows a phase transition. If the quark confinement 
is a consequence of absence of a phase transition for the Yang-Mills theory, 
then the fluctuations of the heat bath must destroy the long-range order of 
the non-linear $\sigma$-model. In order to clarify whether this is true, 
we replace the fluctuations of the heat bath with Gaussian random variables, 
and obtain a Langevin equation which yields the effective action of 
the non-linear $\sigma$-model through analyzing the Fokker-Planck equation. 
It turns out that the fluctuations indeed destroy the long-range order 
of the non-linear $\sigma$-model within a mean field approximation 
estimating a critical point, whereas for the corresponding U(1) gauge theory, 
the phase transition to the massless phase remains against the fluctuations.  
\par
\vfil}\newpage
\Section{Introduction}

We study Euclidean SU(2) Yang-Mills theory on the hypercubic lattice $\ze^d$ 
with dimension $d\ge 4$. It is widely believed that\footnote{See, for example, 
the book \cite{MM}.} 
the gauge theory shows a quark confinement phase with a mass gap 
for all the values of the coupling in dimensions $d=4$. 
On the other hand, the corresponding U(1) gauge theory in dimensions $d=4$ is proven to show 
the existence of a deconfining transition to a massless phase \cite{Guth,FS}. 
Thus it is expected that there exists a crucial difference between 
SU(2) and U(1) gauge theories. 

In this paper, we explore the origin of this difference. 
For this purpose, we go back to the paper by Durhuus and Fr\"ohlich \cite{DF}. 
They showed that the $d$-dimensional Yang-Mills system can be interpreted as  
many $(d-1)$-dimensional non-linear $\sigma$-models which are stacked up 
in the $d$-th direction and coupled through 
$(d-1)$-dimensional external Yang-Mills fields.\footnote{See also related articles 
\cite{Orland,Orland2}.} 
When we give our eye to one of the $(d-1)$-dimensional non-linear $\sigma$-models, 
the system can be interpreted as a $(d-1)$-dimensional non-linear $\sigma$-model 
in a $d$-dimensional heat bath. 
When we turn off the interaction between the non-linear $\sigma$-model and the heat bath, 
the non-linear $\sigma$-model becomes the standard O(4) non-linear $\sigma$-model 
because the gauge group SU(2) is homeomorphic to $3$-sphere $\sp^3$. 
As is well known, the O(4) non-linear $\sigma$-model  
is proven to show a phase transition \cite{FSS} in dimensions greater than or 
equal to three. 
This implies that, if the quark confinement is a consequence of absence 
of a phase transition for the Yang-Mills theory, 
then the fluctuations of the external Yang-Mills fields must destroy 
the long-range order of the O(4) non-linear $\sigma$-model.  

The effective action of the $(d-1)$-dimensional non-linear $\sigma$-model 
can be derived by integrating out the degrees of freedom of the heat bath. 
However, carrying out the integration is very difficult. Instead of doing so, 
we replace the fluctuations of the external Yang-Mills fields with 
Gaussian random variables. Within this approximation, 
the spins of the non-linear $\sigma$-model can be interpreted as ``particles" which move 
on $\sp^3$, acted by the two-body interaction and the random forces.  
Namely the dynamics of the ``particles" obeys a Langevin equation \cite{PW}. 
As is well known, a Langevin dynamics yields a Fokker-Planck equation 
which describes the time evolution of the distribution of the ``particles". 
In the present system, the effective action of the non-linear $\sigma$-model 
can be derived from the steady state solution to the corresponding 
Fokker-Planck equation.  
In the effective action so obtained, the attractive potential between the two ``particles"  
is modified by the fluctuations of the external Yang-Mills fields.  

We show that the height and the width of the barrier of the attractive potential 
depend on the coupling constant of the Yang-Mills theory. 
Roughly speaking, the critical value of the coupling constant 
for the phase transition to a massless phase 
can be estimated by the height and the width of the barrier of the attractive potential. 
Therefore the critical value becomes a function of the coupling constant.  
In consequence, we obtain that within a certain mean field approximation, 
the critical value is always strictly less than the value of the coupling constant 
itself for weak couplings. This implies that the critical value must be equal to zero, 
i.e., there is no phase transition to a massless phase for non-zero coupling constants. 

On the other hand, the corresponding U(1) gauge theory shows that 
the attractive potential does not depend on the coupling constant for weak coupling 
constants within the same approximation. 
Namely the fluctuations of the external Yang-Mills fields does not affect 
the critical behavior of the O(2) non-linear $\sigma$-model. 

This paper is organized as follows. In the next section, we express 
SU(2) Yang-Mills theory in the form of the O(4) non-linear $\sigma$-model 
with a large heat bath, following Durhuus and Fr\"ohlich \cite{DF}. 
In Section~\ref{Sec:Langevin}, we obtain the Langevin equation 
for the ``particles" moving on $\sp^3$, by replacing the fluctuations of the heat bath 
with Gaussian random variables. 
In the standard procedure, the Langevin equation yields the Fokker-Planck equation 
for the distribution of the ``particles".  
In Section~\ref{Sec:SSSFP}, a steady state solution to the Fokker-Planck equation 
is obtained. The result immediately yields the effective action of 
the non-linear $\sigma$-model. Further, we show that the phase transition of 
the O(4) non-linear $\sigma$-model disappears, owing to the fluctuations, 
within a mean field approximation for the effective action so obtained. 
In Section~\ref{Sec:U(1)case}, we apply the same method to the corresponding U(1) 
gauge theory, and show that the phase transition to the massless phase 
remains against the fluctuations. 
 
\Section{Yang-Mills theory as a $\sigma$-model in a heat bath}

Let $\Lambda$ be a sublattice of $\ze^d$. 
The SU(2) gauge field on $\Lambda$ is a map from the oriented 
links or nearest neighbour pairs $\langle {\bf q},{\bf q}'\rangle$ of 
sites, ${\bf q},{\bf q}'$, of the lattice $\Lambda$ into the Lie group $G=$SU(2),  
\begin{equation}
\langle {\bf q},{\bf q}'\rangle\longmapsto U_{{\bf q}{\bf q}'}\in G,
\end{equation}
obeying 
\begin{equation}
U_{{\bf q}'{\bf q}}=\left(U_{{\bf q}{\bf q}'}\right)^{-1}.
\end{equation}
Let $\gamma$ be an oriented path which is written 
$\gamma=\langle {\bf q}_1,{\bf q}_2\rangle \langle {\bf q}_2,{\bf q}_3\rangle\cdots 
\langle {\bf q}_{n-1},{\bf q}_n\rangle$ 
with the oriented links, $\langle {\bf q}_i,{\bf q}_{i+1}\rangle$ of 
the neighboring sites, ${\bf q}_i,{\bf q}_{i+1}$, 
for $i=1,2,\ldots,n-1$. When ${\bf q}_1={\bf q}_n$, the path $\gamma$ is a loop. 
For an oriented path $\gamma$, we write 
\begin{equation}
U_\gamma=U_{{\bf q}_1{\bf q}_2}U_{{\bf q}_2{\bf q}_3}\cdots U_{{\bf q}_{n-1}{\bf q}_n}. 
\end{equation}
The Euclidean action of pure Yang-Mills theory on the lattice 
$\Lambda\subset\ze^d$ is given by  
\begin{equation}
{\cal A}_d^{\rm YM}(\Lambda):=-\frac{1}{2}\sum_{p\subset\Lambda}
{\rm Re}\, {\rm Tr}\, U_{\partial p},
\end{equation}
where $p$ denotes an oriented plaquette(unit square) of $\Lambda$, 
and $\partial p$ is the oriented loop formed by the four sides of $p$. 
The orientation of the loop $\partial p$ obeys the orientation of the plaquette $p$.  
The expectation value is given by 
\begin{equation}
\left\langle\cdots\right\rangle_\Lambda:=Z_\Lambda^{-1}
\int\prod_{b\subset\Lambda}dU_b(\cdots)\exp\left[-\beta{\cal A}_d^{\rm YM}(\Lambda)\right]
\end{equation}
with the inverse temperature $\beta$ and the normalization $Z_\Lambda$, 
where $b$ is a link in $\Lambda$ and $dU_b$ is 
the Haar measure of the gauge group $G=$SU(2). 

Following Durhuus and Fr\"ohlich \cite{DF}, we use the relation 
between the $d$-dimensional Yang-Mills action and a $(d-1)$-dimensional 
non-linear $\sigma$-model. 
The coordinates of a lattice site ${\bf q}$ are denoted  
$(x^{(1)},x^{(2)},\ldots,x^{(d-1)},x^{(d)})=({\bf i},x^{(d)})$ 
with ${\bf i}=(x^{(1)},\ldots,x^{(d-1)})\in\ze^{d-1}$. 
Write $\Lambda_\tau=\Lambda\cap\{{\bf q}:x^{(d)}=\tau\}$ 
for the $(d-1)$-dimensional hyperplane, 
and $\Lambda^0=\Lambda\cap\ze^{d-1}\times\{0\}$ for the projection onto $\ze^{d-1}$ lattice.
Let $U_{{\bf i}{\bf j}}^h(\tau)$ denote the gauge field $U_{{\bf q}{\bf q}'}$ assigned to 
the link $\langle {\bf q},{\bf q}'\rangle$ in $\Lambda_\tau$ 
with ${\bf q}=({\bf i},\tau)$ and ${\bf q}'=({\bf j},\tau)$, 
and $U_{\bf i}^v(\tau)$ the gauge field $U_{{\bf q}{\bf q}'}$ with 
${\bf q}=({\bf i},\tau)$ and ${\bf q}'=({\bf i},\tau+1)$. 
The former are called horizontal gauge fields 
localized at $x^{(d)}=\tau$, and the latter are called vertical gauge fields localized 
in the slice $[\tau,\tau+1]$. Now the Yang-Mills action can be rewritten as 
\begin{equation}
{\cal A}_d^{\rm YM}(\Lambda)=-\frac{1}{2}\sum_\tau\sum_{p\subset\Lambda_\tau}
{\rm Re}\,{\rm Tr}\,U_{\partial p}^h
-\frac{1}{2}\sum_\tau\sum_{\langle {\bf i},{\bf j}\rangle\subset\Lambda^0}
{\rm Re}\,{\rm Tr}\,{U_{\bf i}^v(\tau)}^{-1}U_{{\bf i}{\bf j}}^h(\tau)
U_{\bf j}^v(\tau)U_{{\bf j}{\bf i}}^h(\tau+1).
\end{equation}
The first term in the right-hand side is a sum of Yang-Mills actions which depend on 
the horizontal gauge fields in $(d-1)$-dimensional hyperplane at $x^{(d)}=\tau$. 
As to the second term, the vertical gauge fields in different slices are not 
coupled to each other. Therefore the summand about $\tau$ in the second term is written in 
an action of a $(d-1)$-dimensional non-linear $\sigma$-model for the vertical gauge 
fields as 
\begin{equation}
{\cal A}_{d-1}^\sigma(\Lambda^0;U^h(\tau),U^h(\tau+1))=
-\frac{1}{2}\sum_{\langle {\bf i},{\bf j}\rangle\subset\Lambda^0}
{\rm Re}\,{\rm Tr}\,{U_{\bf i}^v(\tau)}^{-1}U_{{\bf i}{\bf j}}^h(\tau)
U_{\bf j}^v(\tau)U_{{\bf j}{\bf i}}^h(\tau+1)
\label{actionNLS} 
\end{equation}
in the external gauge fields, $U^h(\tau)=\{U_{{\bf i}{\bf j}}^h(\tau)\}$ and 
$U^h(\tau+1)=\{U_{{\bf i}{\bf j}}^h(\tau+1)\}$. 

Let $\sp^3$ denote the $3$-sphere. In order to express the gauge fields 
in terms of spins ${\bf S}\in\sp^3$, 
we use the homeomorphism $\varphi:\sp^3\rightarrow {\rm SU(2)}$ 
which is defined by \cite{DF} 
\begin{equation}
\varphi({\bf S})=
\varphi\left(S^{(0)},S^{(1)},S^{(2)},S^{(3)}\right)=
\left(\matrix{S^{(0)}+iS^{(3)} & -S^{(1)}+iS^{(2)} \cr 
S^{(1)}+iS^{(2)} & S^{(0)}-iS^{(3)} \cr}\right) 
\end{equation}
with the radius $(S^{(0)})^2+(S^{(1)})^2+(S^{(2)})^2+(S^{(3)})^2=1$. 
Then the interaction potential $V_{12}$ between two spins ${\bf S}_1$ and ${\bf S}_2$ 
in the non-linear $\sigma$-model (\ref{actionNLS}) can be written  
\begin{equation}
V_{12}=-\frac{1}{2}\,{\rm Re}\,{\rm Tr}\,\varphi\left({\bf S}_1\right)^{-1}
\varphi\left(\mbox{\boldmath $\sigma$}_1\right)
\varphi\left({\bf S}_2\right)\varphi\left(\mbox{\boldmath $\sigma$}_2\right)^{-1}, 
\end{equation}
where we have written $\mbox{\boldmath $\sigma$}_1$ 
and $\mbox{\boldmath $\sigma$}_2$ for the external horizontal gauge fields.  
When the external gauge fields, $\mbox{\boldmath $\sigma$}_\ell$, take 
the vacuum configurations, 
$\mbox{\boldmath $\sigma$}_1=\mbox{\boldmath $\sigma$}_2=(1,0,0,0)$, 
the interaction becomes that of the O(4) non-linear $\sigma$-model in 
$(d-1)$ dimensions as 
\begin{equation}
V_{12}=-\frac{1}{2}\,{\rm Re}\,{\rm Tr}\,\varphi\left({\bf S}_1\right)^{-1}
\varphi\left({\bf S}_2\right)=-{\bf S}_1\cdot{\bf S}_2=-\sum_{k=0}^3 S_1^{(k)}S_2^{(k)}. 
\end{equation}
As is well known, the O(4) non-linear $\sigma$-model shows 
a long-range order of spins at low temperatures in three or higher dimensions \cite{FSS}. 
The long-range order leads to the perimeter law of the decay of the Wilson loop \cite{DF}. 
The perimeter law implies deconfinement of quarks. 
If the confinement of quarks indeed occurs in the SU(2) gauge theory, 
the fluctuations of the external gauge fields around the vacuum 
must destroy the long-range order of the O(4) non-linear $\sigma$-model. 

In order to take account of the fluctuations around 
the vacuum configuration of the external gauge fields, 
we approximate $\mbox{\boldmath $\sigma$}_\ell$ as  
\begin{equation}
\mbox{\boldmath $\sigma$}_\ell=
\left(\sqrt{1-\left|\hat{\mbox{\boldmath $\sigma$}}_\ell\right|^2},
\hat{\mbox{\boldmath $\sigma$}}_\ell\right)
\approx\left(1,\hat{\mbox{\boldmath $\sigma$}}_\ell\right) 
\end{equation}
with small fluctuations, 
\begin{equation}
\hat{\mbox{\boldmath $\sigma$}}_\ell=\left(\sigma_\ell^{(1)},\sigma_\ell^{(2)},
\sigma_\ell^{(3)}\right),\quad\mbox{for}\ \ell=1,2. 
\end{equation}
We write $\delta\mbox{\boldmath $\sigma$}_\ell=(0,\hat{\mbox{\boldmath $\sigma$}}_\ell)$. 
Then the two-body potential is written 
\begin{equation}
V_{12}\approx -{\bf S}_1\cdot{\bf S}_2-\frac{1}{2}\,{\rm Re}\,{\rm Tr}\, 
\varphi({\bf S}_1)^{-1}\varphi'(\delta\mbox{\boldmath $\sigma$}_1)
\varphi({\bf S}_2)-\frac{1}{2}
\ {\rm Re}\ {\rm Tr}\ 
\varphi({\bf S}_1)^{-1}\varphi({\bf S}_2)\varphi'(-\delta\mbox{\boldmath $\sigma$}_2),
\label{linearV12}
\end{equation}
dropping the second order\footnote{The contributions of the second order of 
the fluctuations $\delta\mbox{\boldmath $\sigma$}_\ell$ give order of 
temperature $T=\beta^{-1}$ in the potential $V_{12}$. Therefore one can expect that 
the contributions of the second order slightly modifies the coupling constants of 
the interaction potentials at low temperatures.} 
in the fluctuations $\delta\mbox{\boldmath $\sigma$}_\ell$. Here we have written 
\begin{equation}
\varphi'(\delta\mbox{\boldmath $\sigma$})
=\left(\matrix{i\sigma^{(3)} & -\sigma^{(1)}+i\sigma^{(2)} \cr 
\sigma^{(1)}+i\sigma^{(2)} & -i\sigma^{(3)} \cr}\right).
\end{equation}
The right-hand side of (\ref{linearV12}) can be written 
\begin{equation}
V_{12}\approx V_0+V_{\rm R}
\end{equation}
with 
\begin{equation}
V_0=-{\bf S}_1\cdot{\bf S}_2
\end{equation}
and
\begin{equation} 
V_{\rm R}=-\sqrt{2}\,\hat{\mbox{\boldmath $\sigma$}}_+\cdot
\left(\hat{\bf S}_1\times\hat{\bf S}_2\right)
-\sqrt{2}\,\hat{\mbox{\boldmath $\sigma$}}_-\cdot
\left(S_1^{(0)}\hat{\bf S}_2-S_2^{(0)}\hat{\bf S}_1\right),
\label{VR0}
\end{equation}
where
\begin{equation}
\hat{\mbox{\boldmath $\sigma$}}_\pm=\frac{1}{\sqrt{2}}
\left(\hat{\mbox{\boldmath $\sigma$}}_2\pm\hat{\mbox{\boldmath $\sigma$}}_1\right),
\end{equation}
and
\begin{equation}
\hat{\bf S}_\ell=\left(S_\ell^{(1)},S_\ell^{(2)},S_\ell^{(3)}\right),
\quad\ell=1,2.
\end{equation}
Thus the present system can be expressed as the O(4) non-linear $\sigma$-model 
in the heat bath. The interaction between the non-linear $\sigma$-model and 
the heat bath is given by $V_{\rm R}$.

\Section{Langevin dynamics for two particles on $\sp^3$.}
\label{Sec:Langevin}

If we can integrate out the degrees of freedom of the heat bath, 
then we can obtain the effective action of the non-linear $\sigma$-model. 
However, it is very difficult problem. 
Instead of this way, we replace the fluctuations of the external gauge fields 
with Gaussian random variables. Then, the spins of the $\sigma$-model 
can be interpreted as the ``particles" which move on $\sp^3$, 
acted by the two-body interaction and the random forces.  

In order to derive the effective two-body interaction between two spins of 
the $\sigma$-model within this approximation, 
we first introduce the Langevin equation for the two ``particles". 
We write ${\hat x}_\ell=(x_\ell^{(1)},x_\ell^{(2)},x_\ell^{(3)})$, $\ell=1,2$, for 
the local coordinates of the two 3-spheres $\sp^3$. 
Then the Langevin equation \cite{PW} is given by 
\begin{equation}
\frac{d}{dt}x_\ell^{(i)}=F_{0,\ell}^{(i)}+F_{{\rm R},\ell}^{(i)},
\quad\ell=1,2;\ \ i=1,2,3.
\label{Langevin}
\end{equation}
with the forces, $F_{0,\ell}^{(i)},F_{{\rm R},\ell}^{(i)}$, which are 
given by the gradient\footnote{See, for example, the book \cite{Sakai}.} 
of the potentials as 
\begin{equation}
F_{0,\ell}^{(i)}=-g^{ij}_{\ \ell}\partial_{j,\ell}V_0
\end{equation}
and 
\begin{equation}
F_{{\rm R},\ell}^{(i)}=-g^{ij}_{\ \ell}\partial_{j,\ell}V_{\rm R},
\end{equation}
where $g^{ij}_{\ \ell}$ is the matrix inverse of the metric tensor $g_{ij,\ell}$ 
for the ``particle" $\ell$, and we have used the Einstein summation convention 
and written 
\begin{equation}
\partial_{i,\ell}=\frac{\partial}{\partial x_\ell^{(i)}}.
\end{equation}

Let $\rho_t({\hat x}_1,{\hat x}_2)$ be the distribution of the two ``particles" 
on $\sp^3\times \sp^3$. 
The expectation value of the function $f({\hat x}_1,{\hat x}_2)$ on $\sp^3\times \sp^3$ 
at time $t$ is given by 
\begin{equation}
\left\langle f\right\rangle_t:=
\int_{\sp^3\times \sp^3}f({\hat x}_1,{\hat x}_2)\rho_t({\hat x}_1,{\hat x}_2)d\mu_1d\mu_2,
\end{equation}
where we have written 
\begin{equation}
d\mu_\ell=\sqrt{{\rm det}\, g_\ell}\, 
dx_\ell^{(1)}dx_\ell^{(2)}dx_\ell^{(3)}\quad\mbox{for \ \ }\ell=1,2.
\end{equation}
For a small $\Delta t>0$, the following relation must hold:
\begin{equation}
\left\langle f\right\rangle_{t+\Delta t}=
\av\int_{\sp^3\times \sp^3}f({\hat x}_1(t+\Delta t),{\hat x}_2(t+\Delta t))
\rho_t({\hat x}_1,{\hat x}_2)d\mu_1d\mu_2 
+{\cal O}((\Delta t)^2),
\label{conservation}
\end{equation}
where $\av$ stands for the average over 
the fluctuations $\hat{\mbox{\boldmath $\sigma$}}_\ell$, $\ell=1,2$, 
and ${\hat x}_\ell(t+\Delta t)$ is the solution of the Langevin 
equation (\ref{Langevin}) with the initial conditions ${\hat x}_\ell(t)={\hat x}_\ell$ 
at time $t$. 
As usual, we assume that, for the short interval $[t,t+\Delta t]$, 
the fluctuations ${\hat \sigma}_\ell^{(i)}$ are constant, and satisfy  
\begin{equation}
\av\left[\sigma_\ell^{(i)}\right]=0,\quad
\av\left[\sigma_\ell^{(i)}\sigma_\ell^{(j)}\right]
=\frac{\alpha}{\Delta t}\delta^{ij}\quad\mbox{and}\quad
\av\left[\sigma_1^{(i)}\sigma_2^{(j)}\right]
=\frac{\alpha'}{\Delta t}\delta^{ij},
\label{variance} 
\end{equation}
where $\alpha$ and $\alpha'$ are a nonnegative constant, and 
$\delta^{ij}$ is the Kronecker delta. Physically, a natural assumption is that 
$\alpha$ and $\alpha'$ satisfy the condition $\alpha>\alpha'>0$. 
{From} the relation between the fluctuations and the temperature of 
the heat bath, both of $\alpha$ and $\alpha'$ are proportional to the temperature $\beta^{-1}$ of 
the heat bath. 

{From} the Langevin equation (\ref{Langevin}), we have 
\begin{equation}
x_\ell^{(i)}(s)-x_\ell^{(i)}(t)=\int_t^sdt'\frac{dx_\ell^{(i)}(t')}{dt}=
\int_t^sdt'F_\ell^{(i)}({\tilde x}(t')),
\end{equation}
where we have written $F_\ell^{(i)}=F_{0,\ell}^{(i)}+F_{R,\ell}^{(i)}$ and 
${\tilde x}(t)=({\hat x}_1(t),{\hat x}_2(t))$.
Using this relation, we obtain
\begin{equation}
F_\ell^{(i)}({\tilde x}(t'))=F_\ell^{(i)}({\tilde x}(t))
+\sum_{m,k}\frac{\partial F_\ell^{(i)}({\tilde x}(t))}{\partial x_m^{(k)}}
\int_t^{t'}dt''F_m^{(k)}({\tilde x}(t''))+\cdots.
\end{equation}
Combining these, the expansion with respect to $\Delta t$ is derived as 
\begin{equation}
x_\ell^{(i)}(t+\Delta t)=x_\ell^{(i)}(t)
+F_\ell^{(i)}({\tilde x}(t))\Delta t+\frac{1}{2}
\sum_{m,k}\frac{\partial F_\ell^{(i)}({\tilde x}(t))
}{\partial x_m^{(k)}}F_m^{(k)}({\tilde x}(t))(\Delta t)^2+\cdots. 
\end{equation}
Substituting this into (\ref{conservation}) and using (\ref{variance}), 
the order of $\Delta t$ yields 
\begin{eqnarray}
\int_Md\mu
f({\tilde x})\frac{\partial \rho_t({\tilde x})}{\partial t} 
&=&\int_Md\mu
\sum_{\ell,i}\frac{\partial f({\tilde x})}{\partial x_\ell^{(i)}}
F_{0,\ell}^{(i)}({\tilde x})\rho_t({\tilde x})\ret
&+&\frac{\Delta t}{2}\int_Md\mu
\sum_{\ell,i;m,j}
\frac{\partial^2 f({\tilde x})}{\partial x_\ell^{(i)}\partial x_m^{(j)}}
\av\left[F_{{\rm R},\ell}^{(i)}({\tilde x})F_{{\rm R},m}^{(j)}({\tilde x})\right]
\rho_t({\tilde x})\ret
&+&\frac{\Delta t}{2}\int_Md\mu
\sum_{\ell,i;n,k}\frac{\partial f({\tilde x})}{\partial x_\ell^{(i)}}
\av\left[\frac{\partial F_{{\rm R},\ell}^{(i)}({\tilde x})}{\partial x_n^{(k)}}
F_{{\rm R},n}^{(k)}({\tilde x})\right]
\rho_t({\tilde x}),
\label{conseDeltat}
\end{eqnarray}
where we have written $M=\sp^3\times\sp^3$ and $d\mu=d\mu_1d\mu_2$. 
Since this equation holds for any function $f$, we can derive 
the equation of the time evolution for the distribution $\rho_t$, 
i.e., the Fokker-Planck equation. 

To this end, consider first the first term in the right-hand side 
of (\ref{conseDeltat}). Note that   
\begin{eqnarray}
\sum_i\frac{\partial f({\tilde x})}{\partial x_\ell^{(i)}}
F_{0,\ell}^{(i)}({\tilde x})\rho_t({\tilde x})
&=&\sum_i\frac{1}{\sqrt{{\rm det}\, g_\ell}}
\frac{\partial}{\partial x_\ell^{(i)}} 
\sqrt{{\rm det}\, g_\ell}\,F_{0,\ell}^{(i)}({\tilde x})f({\tilde x})\rho_t({\tilde x})\ret
&-&\sum_if({\tilde x})\frac{1}{\sqrt{{\rm det}\, g_\ell}}
\frac{\partial}{\partial x_\ell^{(i)}} 
\sqrt{{\rm det}\, g_\ell}\,F_{0,\ell}^{(i)}({\tilde x})\rho_t({\tilde x})\ret
&=&{\rm div}_\ell\left[F_{0,\ell}({\tilde x})f({\tilde x})\rho_t({\tilde x})\right]
-f({\tilde x})\,{\rm div}_\ell\left[F_{0,\ell}({\tilde x})\rho_t({\tilde x})\right],
\end{eqnarray}
where ${\rm div}_\ell$ stands for the divergence for the ``particle" $\ell$. 
Combining this with 
the divergence theorem,\footnote{See, for example, Theorem~5.11 in Chap.~II of 
the book \cite{Sakai}.}  
\begin{equation}
\int_{\sp^3}d\mu_\ell\;{\rm div}_\ell\, v_\ell=0,
\label{GreenTh}
\end{equation}
for a vector field $v_\ell$ on $\sp^3$, 
the first term in the right-hand side of (\ref{conseDeltat}) is written as 
\begin{equation}
\sum_{\ell,i}\int_Md\mu\left(\partial_{i,\ell}f\right)
F_{0,\ell}^{(i)}\rho_t
=-\sum_\ell\int_Md\mu\,f\,{\rm div}_\ell\left(F_{0,\ell}\rho_t\right).
\label{conseF0}
\end{equation}

As to the second and third terms in the right-hand side 
of (\ref{conseDeltat}), we must compute the second moments of the random forces. 
But one can treat these terms in the same way as in the above. 
The detail is given in Appendix~\ref{DerivFPeq}. 
As a result, the Fokker-Planck equation is given by 
\begin{eqnarray}
\frac{\partial \rho_t}{\partial t}&=&-\sum_\ell
{\rm div}_\ell\left(F_{0,\ell}\rho_t\right)
+(\alpha+\alpha')\sum_\ell\left\{\Delta_\ell\rho_t-{\rm div}_\ell
\left[\xi_\ell\,{\rm div}_\ell(\xi_\ell\rho_t)\right]\right\}\ret
& &-(\alpha+\alpha')\left\{{\rm div}_1\left[\mbox{\boldmath $\eta$}_1W
\cdot{\rm div}_2(\mbox{\boldmath $\eta$}_2\rho_t)\right]+
{\rm div}_2\left[\mbox{\boldmath $\eta$}_2W
\cdot{\rm div}_1(\mbox{\boldmath $\eta$}_1\rho_t)\right]\right\}\ret
& &-2\alpha'\sum_{m,n}{\rm div}_m\left[\hat{\mbox{\boldmath $\zeta$}}_m
\cdot{\rm div}_n(\hat{\mbox{\boldmath $\zeta$}}_n\rho_t)\right],
\label{FPeqtot}
\end{eqnarray}
where $\Delta_\ell$ is the Laplacian for the ``particle" $\ell$, and 
we have written $W={\bf S}_1\cdot{\bf S}_2$; 
the vector fields, $\xi_\ell$, $\mbox{\boldmath $\eta$}_\ell$ 
and $\hat{\mbox{\boldmath $\zeta$}}_\ell$, are given by   
\begin{equation}
\xi_\ell^{i}:=g^{ij}_{\ \ell}\partial_{j,\ell}W,
\label{xiell}
\end{equation}
\begin{equation}
\mbox{\boldmath $\eta$}_\ell^i:=g^{ij}_{\ \ell}\partial_{j,\ell}{\bf S}_\ell
\label{eta}
\end{equation}
and 
\begin{equation}
\hat{\mbox{\boldmath $\zeta$}}_\ell^i:=
g^{ij}_{\ \ell}\partial_{j,\ell}
\left(S_1^{(0)}{\hat{\bf S}}_2-S_2^{(0)}{\hat{\bf S}}_1\right)
\label{zeta} 
\end{equation}
for $i=1,2,3$ and $\ell=1,2$. Here the vectors $\mbox{\boldmath $\eta$}_\ell^i$ have 
four components like ${\bf S}_\ell$, 
and $\hat{\mbox{\boldmath $\zeta$}}_\ell^i$ have three components 
like ${\hat{\bf S}}_\ell$. This Fokker-Planck equation can be written 
\begin{equation}
\frac{\partial \rho_t}{\partial t}=-{\rm div}\, J
\quad\mbox{with}\ \ {\rm div}\,J={\rm div}_1J_1+{\rm div}_2J_2
\label{FPeqdivJ}
\end{equation}
in terms of the current $J=(J_1,J_2)$ which is given by 
\begin{equation}
J_{\ \ell}^i=g^{ij}_{\ \ell}J_{j,\ell}
\end{equation}
with 
\begin{eqnarray}
J_{j,1}&=&-(\partial_{j,1}V_0)\rho_t-(\alpha+\alpha')\left\{\partial_{j,1}\rho_t
-\left[(\partial_{j,1}W){\rm div}_1(\xi_1\rho_t)
+W(\partial_{j,1}{\bf S}_1)\cdot{\rm div}_2(\mbox{\boldmath $\eta$}_2\rho_t)\right]
\right\}\ret
& &+2\alpha'\hat{\mbox{\boldmath $\zeta$}}_{j,1}\cdot
\left[{\rm div}_1(\hat{\mbox{\boldmath $\zeta$}}_1\rho_t)
+{\rm div}_2(\hat{\mbox{\boldmath $\zeta$}}_2\rho_t)\right]
\label{current} 
\end{eqnarray}
and with $J_{j,2}$ given by exchanging the subscripts 1 and 2 in $J_{j,1}$. 
Here we have written 
\begin{equation}
\hat{\mbox{\boldmath $\zeta$}}_{i,\ell}:=
\partial_{i,\ell}
\left(S_1^{(0)}{\hat{\bf S}}_2-S_2^{(0)}{\hat{\bf S}}_1\right). 
\end{equation}

\Section{A steady state for the Fokker-Planck dynamics}
\label{Sec:SSSFP}

The effective potential $V_{\rm eff}$ between the two ``particles" is 
derived from a steady distribution $\rho_t=\rho$ 
for the Fokker-Planck equation (\ref{FPeqdivJ}), as in (\ref{steadyrho}) below. 
For a steady distribution $\rho_t=\rho$, 
the Fokker-Planck equation (\ref{FPeqdivJ}) becomes 
${\rm div}\,J=0$. In order to obtain the solution 
near the north pole, ${\bf S}_\ell=(1,0,0,0)$, for $\ell=1,2$, 
we introduce the local coordinates, $(x_\ell,y_\ell,z_\ell)$ for $\ell=1,2$, as  
\begin{equation}
{\bf S}_\ell=\left(\sqrt{1-x_\ell^2-y_\ell^2-z_\ell^2},x_\ell,y_\ell,z_\ell\right).
\end{equation}
We write 
\begin{equation}
{\bf r}=(x,y,z)=(x_1-x_2,y_1-y_2,z_1-z_2)
\end{equation}
and 
\begin{equation}
{\bf R}=(X,Y,Z)=(x_1+x_2,y_1+y_2,z_1+z_2).
\end{equation}
We also write $r=|{\bf r}|$ and $R=|{\bf R}|$. 
In order to solve the partial differential equation ${\rm div}\, J=0$, 
we employ the Cauchy-Kowalevski type expansion\footnote{See, for example, 
Sec.~D of Chap.~1 in the book \cite{Folland}.} with respect to 
small $x_\ell,y_\ell,z_\ell$. 

Let us compute the $x$-component $J_{x,1}$ of the current $J_1$ for the particle 1.
Note that 
\begin{equation}
V_0=-{\bf S}_1\cdot{\bf S}_2=-1+\frac{1}{2}r^2
+\frac{1}{8}({\bf r}\cdot{\bf R})^2+\cdots.
\end{equation}
Immediately, 
\begin{equation}
\frac{\partial V_0}{\partial x_1}=x+\frac{1}{4}({\bf r}\cdot{\bf R})x
+\frac{1}{4}({\bf r}\cdot{\bf R})X+\cdots. 
\end{equation}
Therefore, the first term of $J_{x,1}$ of (\ref{current}) becomes  
\begin{equation}
-(\partial_{x,1}V_0)\rho=\left[-x-\frac{1}{4}({\bf r}\cdot{\bf R})x
-\frac{1}{4}({\bf r}\cdot{\bf R})X+\cdots\right]\rho.
\label{DV0rho}
\end{equation}

In order to treat the rest of the terms of $J_{x,1}$, we assume that 
the steady state solution $\rho_t=\rho$ of ${\rm div}\,J=0$ has the form,
\begin{equation}
\rho=\exp[-\beta V_{\rm eff}],
\label{steadyrho}
\end{equation}
where $V_{\rm eff}$ is the effective potential to be determined, and 
$\beta$ is the inverse temperature of the heat bath. Both of $\alpha$ and $\alpha'$ 
are proportional to the temperature $\beta^{-1}$ as mentioned in the preceding section. 
The effective potential $V_{\rm eff}$ must be vanishing for 
${\bf r}=0$ because the two-body potential (\ref{linearV12}) becomes constant 
irrespective of the external fluctuations for ${\bf S}_1={\bf S}_2$.  
{From} this and taking account of the spherical and exchange symmetries, 
we assume that the effective potential $V_{\rm eff}$ can be expended as 
\begin{equation}
V_{\rm eff}=C_{20}r^2+C_{40}r^4+C_{22}r^2R^2+C_{22}'({\bf r}\cdot{\bf R})^2
+\cdots,
\label{Veff} 
\end{equation}
where $C_{20},C_{40},C_{22}$ and $C_{22}'$ are the coefficients to be determined. 
In the following, we take $\alpha$ and $\alpha'$ to be small, 
and ignore the order of $\alpha$ and $\alpha'$.  

For small $x_\ell,y_\ell,z_\ell$, the current $J_{x,1}$ is written  
\begin{eqnarray}
J_{x,1}&=&\left[-x-\frac{1}{4}({\bf r}\cdot{\bf R})x
-\frac{1}{4}({\bf r}\cdot{\bf R})X\right]\rho
-(\alpha-\alpha')\left(\frac{\partial}{\partial x_1}-\frac{\partial}{\partial x_2}\right)
\rho\ret
&+&(\alpha+\alpha')\left[
x\left(x\frac{\partial\rho}{\partial x_1}+y\frac{\partial\rho}{\partial y_1}
+z\frac{\partial\rho}{\partial z_1}\right)
+x\left(x_2\frac{\partial\rho}{\partial x_2}+y_2\frac{\partial\rho}{\partial y_2}
+z_2\frac{\partial\rho}{\partial z_2}\right)
-\frac{r^2}{2}\frac{\partial\rho}{\partial x_2}\right]\ret
&+&2\alpha'
\left[-x_1\left(x\frac{\partial\rho}{\partial x_1}+y\frac{\partial\rho}{\partial y_1}
+z\frac{\partial\rho}{\partial z_1}\right)+
x_2\left(x_1\frac{\partial\rho}{\partial x_1}+y_1\frac{\partial\rho}{\partial y_1}
+z_1\frac{\partial\rho}{\partial z_1}\right)\right.\ret
& &\left.-\left(\frac{3}{2}x+\frac{1}{2}X\right)
\left(x_2\frac{\partial\rho}{\partial x_2}+y_2\frac{\partial\rho}{\partial y_2}
+z_2\frac{\partial\rho}{\partial z_2}\right)-r_2^2\frac{\partial\rho}{\partial x_1}
+\frac{1}{2}(r_1^2+r_2^2)\frac{\partial\rho}{\partial x_2}\right]+\cdots.\ret
\label{Jx1expand}
\end{eqnarray}
The derivation is given in Appendix~\ref{appendix:Jx1expand}. 
Let us substitute $\rho$ of (\ref{steadyrho}) with the effective 
potential (\ref{Veff}) into this right-hand side. 
First of all, since the leading order which is proportional 
to $x\exp[-\beta V_{\rm eff}]$ must be vanishing, 
we have  
\begin{equation}
4\beta(\alpha-\alpha')C_{20}=1.
\end{equation}
Since we can choose 
\begin{equation}
\beta=\frac{1}{\alpha-\alpha'} 
\end{equation}
without loss of generality, we have 
\begin{equation}
C_{20}=\frac{1}{4}.
\end{equation}
Using these, we get 
\begin{equation}
-(\alpha-\alpha')\left(\frac{\partial}{\partial x_1}-\frac{\partial}{\partial x_2}\right)
\exp\left[-\beta V_{\rm eff}\right]
=\left(\frac{\partial V_{\rm eff}}{\partial x_1}-\frac{\partial V_{\rm eff}}
{\partial x_2}\right)\exp\left[-\beta V_{\rm eff}\right]
\end{equation}
with
\begin{equation}
\left(\frac{\partial}{\partial x_1}-\frac{\partial}{\partial x_2}\right)V_{\rm eff}
=x+8C_{40}r^2x+4C_{22}R^2x+4C_{22}'({\bf r}\cdot{\bf R})X+\cdots.
\end{equation}
Moreover we have 
\begin{equation}
\left(x\frac{\partial}{\partial x_1}+y\frac{\partial}{\partial y_1}
+z\frac{\partial}{\partial z_1}\right)\rho=
\left(-\frac{1}{2}\beta r^2+\cdots\right)\exp[-\beta V_{\rm eff}],
\end{equation}
\begin{equation}
\left(x_1\frac{\partial}{\partial x_1}+y_1\frac{\partial}{\partial y_1}
+z_1\frac{\partial}{\partial z_1}\right)\rho=
\left[-\frac{1}{4}\beta r^2-\frac{1}{4}\beta({\bf r}\cdot{\bf R})+\cdots\right]
\exp[-\beta V_{\rm eff}],
\end{equation}
\begin{equation}
\left(x_2\frac{\partial}{\partial x_2}+y_2\frac{\partial}{\partial y_2}
+z_2\frac{\partial}{\partial z_2}\right)\rho=
\left[-\frac{1}{4}\beta r^2+\frac{1}{4}\beta({\bf r}\cdot{\bf R})+\cdots\right]
\exp[-\beta V_{\rm eff}],
\end{equation}
\begin{equation}
-\frac{r^2}{2}\frac{\partial\rho}{\partial x_2}=
\left[-\frac{\beta}{4}xr^2+\cdots\right]\exp[-\beta V_{\rm eff}]
\end{equation}
and 
\begin{equation}
-r_2^2\frac{\partial\rho}{\partial x_1}
+\frac{1}{2}(r_1^2+r_2^2)\frac{\partial\rho}{\partial x_2}=\frac{\beta}{4}x
\left[r^2+R^2-({\bf r}\cdot{\bf R})\right]\exp[-\beta V_{\rm eff}]+\cdots.
\end{equation}
Substituting these into (\ref{Jx1expand}), we obtain 
\begin{eqnarray}
J_{x,1}\exp[\beta V_{\rm eff}]&=&\left[8C_{40}-1\right]r^2x
+\frac{\alpha'\beta}{2}\left[r^2X-({\bf r}\cdot{\bf R})x\right]\ret
&+&\left[4C_{22}+\frac{\alpha'\beta}{2}\right]R^2x
+\left[4C_{22}'-\frac{(\alpha+\alpha')\beta}{4}\right]({\bf r}\cdot{\bf R})X
+\cdots.
\end{eqnarray}
{From} ${\rm div}\,J=0$, the coefficients must satisfy the relations,  
\begin{equation}
5(8C_{40}-1)+\alpha'\beta=0
\end{equation}
and
\begin{equation}
3\left[4C_{22}+\frac{\alpha'\beta}{2}\right]
+\left[4C_{22}'-\frac{(\alpha+\alpha')\beta}{4}\right]=0.
\label{CC'relation}
\end{equation}
Using these relations, the current $J_{x,1}$ can be written
\begin{equation}
J_{x,1}=\left\{-\frac{\alpha'\beta}{5}r^2x+\frac{\alpha'\beta}{2}
\left[r^2X-({\bf r}\cdot{\bf R})x\right]+
A\left[R^2x-3({\bf r}\cdot{\bf R})X\right]\right\}
\exp[-\beta V_{\rm eff}]+\cdots
\end{equation}
with the constant, 
\begin{equation}
A=4C_{22}+\frac{\alpha'\beta}{2},
\end{equation}
which we cannot determine in the present method. 
Clearly one notices that in ${\rm div}\, J$, there appear the other terms, 
\begin{equation}
\frac{1}{5}\alpha'\beta^2 r^4 \quad\mbox{and}\quad
-A\beta[r^2R^2-3({\bf r}\cdot{\bf R})^2].
\label{sterms}
\end{equation} 
These are higher order in powers of the local coordinates but order of $\beta$. 
Since the equation ${\rm div}\,J=0$ must hold, 
this implies that there must exist 
some terms of order of $\beta$ in the effective potential $V_{\rm eff}$ 
so as to cancel the above terms of (\ref{sterms}).   

When both of the coefficients $C_{22}$ and $C_{22}'$ depend on $\beta$, 
the corresponding terms may appear in the expansion. 
In this case, from (\ref{CC'relation}), we have 
\begin{equation}
C_{22}\sim C\beta \quad \mbox{and}\quad C_{22}'\sim -3C\beta
\end{equation}
with some constant $C$ for a large $\beta$. Substituting these into $V_{\rm eff}$, 
we have 
\begin{equation}
V_{\rm eff}\sim \frac{1}{4}r^2+C_{40}r^4
+C\beta[r^2R^2-3({\bf r}\cdot{\bf R})^2].
\end{equation}
This leads to instability of binding of the two particles because 
the value of $R^2$ is expected to become larger than order of $\beta^{-1}$ in 
the thermal equilibrium. 
Thus we require that both of $C_{22}$ and $C_{22}'$ are order of 1. 

In consequence, we need the following terms in the effective potential $V_{\rm eff}$: 
\begin{equation}
C_{60}r^6, \quad C_{42}r^4R^2, \quad C_{42}'r^2({\bf r}\cdot{\bf R})^2. 
\end{equation}
Here all the coefficients, $C_{60}, C_{42}, C_{42}'$, are proportional to $\beta$ for 
a large $\beta$. In the same way as in the above, we can determine these coefficients as 
\begin{equation}
C_{60}=-\frac{3!}{7!}\alpha'\beta^2,\quad
C_{42}=\frac{1}{56}A\beta\quad\mbox{and}\quad
C_{42}'=-\frac{3}{56}A\beta 
\end{equation}
so as to cancel the above terms (\ref{sterms}) which appear in ${\rm div}\,J$. 
As a result, the dominant contributions in the effective potential $V_{\rm eff}$ 
are given by 
\begin{equation}
V_{\rm eff}\sim \frac{1}{4}r^2-\frac{3!}{7!}\alpha'\beta^2r^6
+\frac{1}{56}A\beta r^2[r^2R^2-3({\bf r}\cdot{\bf R})^2]
\label{Vefffinal}
\end{equation}
for a large $\beta$ because the second, third and fourth terms in the right-hand side of 
(\ref{Veff}) do not affect the critical behavior. 

Now we discuss the critical behavior of the $(d-1)$-dimensional 
$\sigma$ model with the above two-body interaction $V_{\rm eff}$. 
Consider first the case of $A=0$. 
Namely the effective potential is given by
\begin{equation}
V_{\rm eff}\sim \frac{1}{4}r^2-\frac{3!}{7!}\alpha'\beta^2r^6 
\end{equation}
for small $r$ and large $\beta$. The second term lowers the potential 
barrier. Within a mean-field approximation \cite{MA}, 
the critical temperature $T_{\rm C}$ can be estimated by 
the volume and the height of the potential well. 
More precisely, $T_{\rm C}\sim({\rm volume})\times({\rm height})$. 
In the present case, the width $w$ and the height $h$ of the effective potential 
$V_{\rm eff}$ are estimated as 
\begin{equation}
w\sim (\lambda\beta)^{-1/4},\quad h\sim (\lambda\beta)^{-1/2}, 
\end{equation}
where we have written 
\begin{equation}
\lambda=12\cdot\frac{3!}{7!}\alpha'\beta.
\end{equation}
Therefore the critical temperature $T_{\rm C}$ is estimated as
\begin{equation}
T_{\rm C}\sim w^3\times h\sim (\lambda\beta)^{-5/4}. 
\end{equation}
This is lower than $\beta^{-1}$ for small temperature $T=\beta^{-1}$. 
This implies that the true critical temperature must be equal to zero. 

In the case of $A\ne 0$, the third term in the right-hand side of 
(\ref{Vefffinal}) may heighten the potential barrier if $R^2$ does not 
take a small value. But it is impossible that the term heightens the potential 
barrier in all the directions of ${\bf r}$.   
Thus we reach the same conclusion, $T_{\rm C}=0$. 

Let us make the following two remarks: 
\begin{enumerate}
\item Our argument can be applied to the systems in arbitrary dimensions. 
Therefore a reader might think that our method suggests no phase transition 
for non-Abelian lattice gauge theory also in five or higher dimensions. 
On this point, we should remark the following: We used the two-body approximation, 
considering only a single plaquette. When dealing with two plaquettes 
within our method, three- and four-body interactions would appear 
in the effective potential for the non-linear $\sigma$-model. 
The resulting interactions may change the conclusion of this section. 
Namely a high-dimensional system may exhibit a phase transition. 
Actually, in five or higher dimensions, the effect of the three- 
or four-body interactions may not be ignored because the number of the neighboring 
plaquettes for a fixed plaquette becomes large, compared to low-dimensional systems. 
However, taking account of such interactions is not so easy.

\item Consider the O(4) non-linear $\sigma$-model on the three-dimensional 
lattice with the effective two-body interaction which we obtained. 
Then the correlation length of the model leads to an estimate of the string tension 
\cite{DF,Orland}.  
Does the scaling limit so obtained give the standard continuum? 
This problem must be very important. But it is very difficult to compute 
the low temperature asymptotics of the correlation length for 
such a weakly attractive potential.     
\end{enumerate}

\Section{Difference between U(1) and SU(2) gauge theories}
\label{Sec:U(1)case}

Let us see difference between U(1) and SU(2) gauge theories. 

For this purpose, we apply the present method to the abelian case $G=$U(1). 
In the case, the gauge field $U_b$ on a link $b$ is written 
\begin{equation}
U_b=\exp[i\theta_b]
\end{equation}
in terms of the angle variable $\theta_b\in[0,2\pi)$. Therefore the two-body 
interaction $V_{12}$ between $\theta_1$ and $\theta_2$ is written 
\begin{equation} 
V_{12}=-\cos(\theta_1-\theta_2+\sigma_1-\sigma_2), 
\end{equation}
where $\sigma_1$ and $\sigma_2$ are the angle variables of the external fields. 
We write $\theta=\theta_1-\theta_2$ and $\delta\sigma=\sigma_1-\sigma_2$, and 
assume that $\delta\sigma$ is a small fluctuation. Under this assumption, 
the potential can be approximated as 
\begin{equation}
V_{12}\approx-\cos\theta+\delta\sigma\sin\theta.
\end{equation} 
Then the Langevin equation is given by 
\begin{equation}
\frac{d\theta}{dt}=-\sin\theta-\delta\sigma\cos\theta.
\label{LangevinU1}
\end{equation}
As usual, we assume 
\begin{equation}
\av[(\delta\sigma)^2]=\frac{\alpha}{\Delta t}
\end{equation}
for a small $\Delta t$. 
In the same way as in the SU(2) case, we obtain the Fokker-Planck equation, 
\begin{equation}
\frac{\partial\rho_t}{\partial t}=
\left[\frac{\partial}{\partial \theta}\sin\theta+
\frac{\alpha}{2}\frac{\partial}{\partial\theta}\sin\theta\cos\theta
+\frac{\alpha}{2}\frac{\partial^2}{\partial\theta^2}\cos^2\theta\right]\rho_t.
\end{equation}
For a steady state $\rho_t=\rho$, we have 
\begin{equation}
\left[\sin\theta+\frac{\alpha}{2}\sin\theta\cos\theta+
\frac{\alpha}{2}\frac{\partial}{\partial\theta}\cos^2\theta\right]\rho=0.
\end{equation}
One can easily find the solution, 
\begin{equation} 
\rho=\cases{\displaystyle{(\cos\theta)^{-1}
\exp\left[-2\alpha^{-1}/{\cos\theta}\right]}, 
& for $-\pi/2<\theta<\pi/2$;\cr
\quad 0, & otherwise.}
\end{equation}
Since the diffusion disappears at $\theta=\pm \pi/2$ in 
the right-hand side of (\ref{LangevinU1}), the ``particle" cannot 
move beyond the points. 
Clearly, we have 
\begin{equation}
\rho\sim{\rm const.}\exp[-\alpha^{-1}\theta^2]
\end{equation}
for a small $\theta$. 
Thus there is no term which is proportional to $\alpha^{-1}$ or 
higher powers of $\alpha^{-1}$ in the effective potential, 
and the critical behavior can be expected to be the same as 
the standard O(2) nonlinear-$\sigma$ model. 
This is consistent with the rigorous result of \cite{Guth,FS}.

\appendix 

\Section{Derivation of the Fokker-Planck equation}
\label{DerivFPeq}

Consider first the case with $\alpha'=0$ in (\ref{variance}).
We introduce $\sigma^{ij}$ satisfying $\sigma^{ji}=-\sigma^{ij}$ with  
\begin{equation}
(\sigma^{01},\sigma^{02},\sigma^{03})
=(\sigma_+^{(1)},\sigma_+^{(2)},\sigma_+^{(3)}),\quad
\mbox{and}\quad
(\sigma^{23},\sigma^{31},\sigma^{12})
=(\sigma_-^{(1)},\sigma_-^{(2)},\sigma_-^{(3)}). 
\end{equation}
Then the random potential $V_{\rm R}$ of (\ref{VR0}) can be written 
\begin{equation}
V_{\rm R}=-\frac{1}{\sqrt{2}}\varepsilon_{ijk\ell}\,\sigma^{ij}\,
S_1^{(k)}S_2^{(\ell)},
\label{VR} 
\end{equation}
where $\varepsilon_{ijk\ell}$ is completely antisymmetric and satisfies 
$\varepsilon_{0123}=+1$, and we have used the Einstein summation convention.  
{From} $\alpha'=0$, we have 
\begin{equation}
\av\left[\sigma^{\alpha\beta}\sigma^{mn}\right]=\frac{\alpha}{\Delta t}
\left(\delta^{\alpha m}\delta^{\beta n}-\delta^{\alpha n}\delta^{\beta m}\right).
\label{variance2}
\end{equation}
Using (\ref{VR}) and (\ref{variance2}), we obtain 
\begin{eqnarray}
& &\av\left[\left(\partial_{\ell,1}V_{\rm R}\right)
\left(\partial_{k,1}V_{\rm R}\right)\right]\ret&=&
\frac{1}{2}\av\left[\varepsilon_{\alpha\beta\gamma\delta}\sigma^{\alpha\beta}
\left(\partial_{\ell,1}S_1^{(\gamma)}\right)S_2^{(\delta)}
\varepsilon_{mnst}\sigma^{mn}\left(\partial_{k,1}S_1^{(s)}\right)S_2^{(t)}\right]\ret
&=&\frac{\alpha}{2\Delta t}\varepsilon_{\alpha\beta\gamma\delta}\varepsilon_{mnst}
(\delta^{\alpha m}\delta^{\beta n}-\delta^{\alpha n}\delta^{\beta m})
\left(\partial_{\ell,1}S_1^{(\gamma)}\right)S_2^{(\delta)}
\left(\partial_{k,1}S_1^{(s)}\right)S_2^{(t)}\ret
&=&\frac{2\alpha}{\Delta t}\sum_{\gamma,\delta}
\left[\left(\partial_{\ell,1}S_1^{(\gamma)}\right)
\left(\partial_{k,1}S_1^{(\gamma)}\right)S_2^{(\delta)}S_2^{(\delta)}
-\left(\partial_{\ell,1}S_1^{(\gamma)}\right)S_2^{(\gamma)}
\left(\partial_{k,1}S_1^{(\delta)}\right)S_2^{(\delta)}\right].
\end{eqnarray}
Using the metric  
\begin{equation}
g_{ij,\ell}=\frac{\partial {\bf S}_\ell}{\partial x_\ell^{(i)}}\cdot
\frac{\partial {\bf S}_\ell}{\partial x_\ell^{(j)}}
\end{equation}
of $\sp^3$ for the ``particle" $\ell$, the above result is written 
\begin{equation}
\av\left[\left(\partial_{\ell,1}V_{\rm R}\right)
\left(\partial_{k,1}V_{\rm R}\right)\right]=
\frac{2\alpha}{\Delta t}\left[g_{\ell k,1}-(\partial_{\ell,1}W)
(\partial_{k,1}W)\right]
\label{avp1VRp1VR}
\end{equation}
and
\begin{equation}
\av\left[\left(\partial_{\ell,2}V_{\rm R}\right)
\left(\partial_{k,2}V_{\rm R}\right)\right]=
\frac{2\alpha}{\Delta t}\left[g_{\ell k,2}-(\partial_{\ell,2}W)
(\partial_{k,2}W)\right],
\end{equation}
where we have written $W={\bf S}_1\cdot{\bf S}_2$. Similarly, we have 
\begin{equation}
\av\left[\left(\partial_{k,1}\partial_{j,1}V_{\rm R}\right)
\left(\partial_{\ell,1}V_{\rm R}\right)\right]
=\frac{2\alpha}{\Delta t}\sum_{\gamma,\delta}
\left[\frac{\partial^2 S_1^{(\gamma)}}{\partial x_1^{(k)}
\partial x_1^{(j)}}\frac{\partial S_1^{(\gamma)}}{\partial x_1^{(\ell)}}
S_2^{(\delta)}S_2^{(\delta)}
-\frac{\partial^2 S_1^{(\gamma)}}{\partial x_1^{(k)}
\partial x_1^{(j)}}S_2^{(\gamma)}
\frac{\partial S_1^{(\delta)}}{\partial x_1^{(\ell)}}
S_2^{(\delta)}\right].
\end{equation}
Combining this with 
\begin{equation}
\sum_\gamma
\frac{\partial^2 S_1^{(\gamma)}}{\partial x_1^{(k)}
\partial x_1^{(j)}}\frac{\partial S_1^{(\gamma)}}{\partial x_1^{(\ell)}}
=\Gamma_{kj,1}^mg_{m\ell,1},
\end{equation}
we obtain 
\begin{equation}
\av\left[\left(\partial_{k,1}\partial_{j,1}V_{\rm R}\right)
\left(\partial_{\ell,1}V_{\rm R}\right)\right]
=\frac{2\alpha}{\Delta t}\left[\Gamma_{kj,1}^mg_{m\ell,1}
-\left(\partial_{k,1}\partial_{j,1}W\right)
\left(\partial_{\ell,1}W\right)\right],
\label{avpp1VRp1VR}
\end{equation}
where $\Gamma_{k\ell,1}^m$ are the Christoffel symbols \cite{Sakai}. 
In the same way, we get 
\begin{equation}
\av\left[\left(\partial_{\ell,1}V_{\rm R}\right)
\left(\partial_{k,2}V_{\rm R}\right)\right]
=-\frac{2\alpha}{\Delta t}W\left(\partial_{\ell,1}\partial_{k,2}W\right)
\label{avp1VRp2VR}
\end{equation}
and
\begin{equation}
\av\left[\left(\partial_{k,2}\partial_{j,1}V_{\rm R}\right)
\left(\partial_{\ell,2}V_{\rm R}\right)\right]
=-\frac{2\alpha}{\Delta t}\left(\partial_{k,2}W\right)
\left(\partial_{j,1}\partial_{\ell,2}W\right).
\label{avp2p1VRp2VR} 
\end{equation}
Using (\ref{avp1VRp1VR}), we have 
\begin{eqnarray}
\av\left[F_{{\rm R},1}^{(i)}F_{{\rm R},1}^{(j)}\right]
&=&\av\left[g^{i\ell}_{\ 1}\left(\partial_{\ell,1}V_{\rm R}\right)
g^{jk}_{\ 1}\left(\partial_{k,1}V_{\rm R}\right)\right]\ret
&=&\frac{2\alpha}{\Delta t}g^{i\ell}_{\ 1}g^{jk}_{\ 1}
\left[g_{\ell k,1}-(\partial_{\ell,1}W)(\partial_{k,1}W)\right]\ret
&=&\frac{2\alpha}{\Delta t}\left(g^{ij}_{\ 1}-\xi_1^i\xi_1^j\right),
\label{avFRFR}
\end{eqnarray}
where $\xi_\ell^{i}$ is the vector field which is given by (\ref{xiell}).  
{From} (\ref{avp1VRp1VR}) and (\ref{avpp1VRp1VR}), we obtain 
\begin{eqnarray}
\sum_k\av\left[\frac{\partial F_{{\rm R},1}^{(i)}}{\partial x_1^k}
F_{{\rm R},1}^{(k)}\right]&=&\av\left[\left(\partial_{k,1}g^{ij}_{\ 1}
\partial_{j,1}V_{\rm R}\right)
\left(g^{k\ell}_{\ 1}\partial_{\ell,1}V_{\rm R}\right)\right]\ret
&=&\left(\partial_{k,1}g^{ij}_{\ 1}\right)g^{k\ell}_{\ 1}\,
\av\left[\left(\partial_{j,1}V_{\rm R}\right)
\left(\partial_{\ell,1}V_{\rm R}\right)\right]
+g^{ij}_{\ 1}g^{k\ell}_{\ 1}\,
\av\left[\left(\partial_{k,1}\partial_{j,1}V_{\rm R}\right)
\left(\partial_{\ell,1}V_{\rm R}\right)\right]\ret
&=&\frac{2\alpha}{\Delta t}\left(\partial_{k,1}g^{ij}_{\ 1}\right)g^{k\ell}_{\ 1}
\left[g_{j\ell,1}-\left(\partial_{j,1}W\right)\left(\partial_{\ell,1}W\right)\right]\ret
&+&\frac{2\alpha}{\Delta t}g^{ij}_{\ 1}g^{k\ell}_{\ 1}
\left[\Gamma_{kj,1}^mg_{m\ell,1}
-\left(\partial_{k,1}\partial_{j,1}W\right)\left(\partial_{\ell,1}W\right)\right]\ret
&=&\frac{2\alpha}{\Delta t}\left[\partial_{j,1}g^{ij}_{\ 1}
+g^{ij}_{\ 1}\Gamma_{kj,1}^k-\left(\partial_{k,1}\xi_1^i\right)\xi_1^k\right]\ret
&=&\frac{2\alpha}{\Delta t}
\left[\frac{1}{\sqrt{{\rm det}\, g_1}}\partial_{j,1}g^{ij}_{\ 1}\sqrt{{\rm det}\,g_1}
-\left(\partial_{k,1}\xi_1^i\right)\xi_1^k\right]
\label{avParFRFR}
\end{eqnarray}
where we have used\footnote{See, for example, Sec.7 of Chap.~I of 
the book \cite{Eisenhart}.}
\begin{equation}
\Gamma_{kj,1}^k=\frac{1}{\sqrt{{\rm det}\,g_1}}\partial_{j,1}\sqrt{{\rm det}\,g_1}.
\end{equation}
In the same way, the relations (\ref{avp1VRp2VR}) and (\ref{avp2p1VRp2VR}) yield 
\begin{eqnarray}
\av\left[F_{{\rm R},1}^{(i)}F_{{\rm R},2}^{(j)}\right]&=&
g^{i\ell}_{\ 1}g^{jk}_{\ 2}\,\av
\left[\left(\partial_{\ell,1}V_{\rm R}\right)
\left(\partial_{k,2}V_{\rm R}\right)\right]\ret
&=&-\frac{2\alpha}{\Delta t}g^{i\ell}_{\ 1}g^{jk}_{\ 2}\,
W\left(\partial_{\ell,1}\partial_{k,2}W\right)
\label{avFR1FR2}
\end{eqnarray}
and 
\begin{eqnarray}
\sum_k\av\left[\frac{\partial F_{{\rm R},1}^{(i)}}{\partial x_2^k}
F_{{\rm R},2}^{(k)}\right]&=&\av\left[\left(\partial_{k,2}g^{ij}_{\ 1}
\partial_{j,1}V_{\rm R}\right)
\left(g^{k\ell}_{\ 2}\partial_{\ell,2}V_{\rm R}\right)\right]\ret
&=&g^{ij}_{\ 1}g^{k\ell}_{\ 2}\,
\av\left[\left(\partial_{k,2}\partial_{j,1}V_{\rm R}\right)
\left(\partial_{\ell,2}V_{\rm R}\right)\right]\ret
&=&-\frac{2\alpha}{\Delta t}g^{ij}_{\ 1}g^{k\ell}_{\ 2}
\left(\partial_{k,2}W\right)\left(\partial_{j,1}\partial_{\ell,2}W\right),
\label{avParFR1FR2}
\end{eqnarray}
respectively. The contribution from the two random forces $F_{{\rm R},\ell}$ 
with the same indexes $\ell=1$ in the right-hand side of (\ref{conseDeltat}) becomes 
\begin{eqnarray}
I_{11}&:=&\frac{\Delta t}{2}\left\{\sum_{i,j}\int_Md\mu\, 
\frac{\partial^2 f}{\partial x_1^{(i)}\partial x_1^{(j)}}
\av\left[F_{{\rm R},1}^{(i)}F_{{\rm R},1}^{(j)}\right]
+\sum_{i,k}\int_Md\mu\,
\frac{\partial f}{\partial x_1^{(i)}}
\av\left[\frac{\partial F_{{\rm R},1}^{(i)}}{\partial x_1^{(k)}}F_{{\rm R},1}^{(k)}
\right]\right\}\rho_t\ret
&=&\alpha\sum_{i,j}\int_Md\mu\,\frac{\partial^2 f}{\partial x_1^{(i)}\partial x_1^{(j)}}
\left(g^{ij}_{\ 1}-\xi_1^i\xi_1^j\right)\rho_t\ret
&+&\alpha\sum_i\int_Md\mu\,\frac{\partial f}{\partial x_1^{(i)}}
\left[\frac{1}{\sqrt{{\rm det}\, g_1}}\partial_{j,1}g^{ij}_{\ 1}
\sqrt{{\rm det}\, g_1}-\left(\partial_{k,1}\xi_1^i\right)\xi_1^k\right]\rho_t,
\label{con11} 
\end{eqnarray}
where we have used (\ref{avFRFR}) and (\ref{avParFRFR}). Note that
\begin{eqnarray}
& &\int_Md\mu\,\left[
\sum_{i,j}g^{ij}_{\ 1}\frac{\partial^2 f}{\partial x_1^{(i)}\partial x_1^{(j)}}
+\sum_i\left(\frac{1}{\sqrt{{\rm det}\, g_1}}\partial_{j,1}g^{ij}_{\ 1}
\sqrt{{\rm det}\, g_1}\right)\frac{\partial f}{\partial x_1^{(i)}}\right]\rho_t\ret
&=&\int_Md\mu\,\left(\frac{1}{\sqrt{{\rm det}\, g_1}}\partial_{j,1}g^{ij}_{\ 1}
\sqrt{{\rm det}\, g_1}{\partial_{i,1} f}\right)\rho_t\ret
&=& \int_Md\mu\,(\Delta_1 f)\rho_t
=\int_Md\mu\,f\left(\Delta_1\rho_t\right),
\label{Laplaccont}
\end{eqnarray}
where the second equality follows from the property\footnote{See, for example, 
Corollary~5.13 in Chap.~II of the book \cite{Sakai}.} of the Laplacian $\Delta_\ell$. 
The rest of the contributions in the right-hand side of (\ref{con11}) are 
computed as 
\begin{eqnarray}
& &\int_Md\mu\,\left[
\sum_{i,j}\frac{\partial^2 f}{\partial x_1^{(i)}\partial x_1^{(j)}}\xi_1^i\xi_1^j
+\sum_i\frac{\partial f}{\partial x_1^{(i)}}
\left(\partial_{k,1}\xi_1^i\right)\xi_1^k\right]\rho_t\ret
&=&\int_Md\mu\,\left[\frac{1}{\sqrt{{\rm det}\, g_1}}\partial_{i,1}
\sqrt{{\rm det}\,g_1}(\partial_{j,1}f)\xi_1^i\xi_1^j\rho_t
-(\partial_{j,1}f)\frac{1}{\sqrt{{\rm det}\, g_1}}\partial_{i,1}
\sqrt{{\rm det}\,g_1}\xi_1^i\xi_1^j\rho_t\right]\ret
&+&\int_Md\mu\,(\partial_{j,1}f)(\partial_{i,1}\xi_1^j)\xi_1^i\rho_t\ret
&=&-\int_Md\mu\,(\partial_{j,1}f)\xi_1^j
\frac{1}{\sqrt{{\rm det}\, g_1}}\partial_{i,1}
\sqrt{{\rm det}\,g_1}\xi_1^i\rho_t\ret
&=&-\int_Md\mu\,(\partial_{j,1}f)\xi_1^j\,
{\rm div}_1\left[\xi_1\rho_t\right]\ret
&=&-\int_Md\mu\,\frac{1}{\sqrt{{\rm det}\, g_1}}\partial_{j,1}
\sqrt{{\rm det}\,g_1}\xi_1^jf\,
{\rm div}_1\left(\xi_1\rho_t\right)\ret
&+&\int_Md\mu\,f\frac{1}{\sqrt{{\rm det}\, g_1}}\partial_{j,1}
\sqrt{{\rm det}\,g_1}\xi_1^j\,
{\rm div}_1\left(\xi_1\rho_t\right)\ret
&=&\int_Md\mu\,f\,{\rm div}_1\left[\xi_1\,
{\rm div}_1\left(\xi_1\rho_t\right)\right],
\end{eqnarray}
where we have used the divergence theorem (\ref{GreenTh}). 
Substituting this and (\ref{Laplaccont}) into (\ref{con11}), we obtain 
\begin{equation}
I_{11}=\alpha\int_Md\mu\, f
\left\{\Delta_1\rho_t-{\rm div}_1\left[\xi_1{\rm div}_1(\xi_1\rho_t)\right]\right\}.
\label{con11f}
\end{equation}

Next consider the contribution from the two random forces $F_{{\rm R},\ell}$ 
with different indexes, $\ell=1$ and $\ell=2$, in the right-hand side 
of (\ref{conseDeltat}). 
Using (\ref{avFR1FR2}) and (\ref{avParFR1FR2}), we obtain  
\begin{eqnarray}
I_{12}&:=&\frac{\Delta t}{2}\left\{\sum_{i,j}\int_Md\mu\, 
\frac{\partial^2 f}{\partial x_1^{(i)}\partial x_2^{(j)}}
\av\left[F_{{\rm R},1}^{(i)}F_{{\rm R},2}^{(j)}\right]
+\sum_{i,k}\int_Md\mu\,
\frac{\partial f}{\partial x_1^{(i)}}
\av\left[\frac{\partial F_{{\rm R},1}^{(i)}}{\partial x_2^{(k)}}F_{{\rm R},2}^{(k)}
\right]\right\}\rho_t\ret
&=&-\alpha\int_Md\mu\,\sum_{i,j} 
\frac{\partial^2 f}{\partial x_1^{(i)}\partial x_2^{(j)}}
g^{i\ell}_{\ 1}g^{jk}_{\ 2}W(\partial_{\ell,1}\partial_{k,2}W)\rho_t\ret
& &-\alpha\int_Md\mu\,\sum_i
\frac{\partial f}{\partial x_1^{(i)}}g^{ij}_{\ 1}g^{k\ell}_{\ 2}
(\partial_{k,2}W)(\partial_{j,1}\partial_{\ell,2}W)\rho_t\ret
&=&-\alpha\int_Md\mu\,\frac{1}{\sqrt{{\rm det}\,g_2}}\partial_{j,2}
\sqrt{{\rm det}\,g_2}g^{jk}_{\ 2}(\partial_{i,1}f)g^{i\ell}_{\ 1}
(\partial_{\ell,1}\partial_{k,2}W)W\rho_t\ret
& &+\alpha\int_Md\mu\,(\partial_{i,1}f)
\frac{1}{\sqrt{{\rm det}\,g_2}}\partial_{j,2}
\sqrt{{\rm det}\,g_2}g^{jk}_{\ 2}g^{i\ell}_{\ 1}
(\partial_{\ell,1}\partial_{k,2}W)W\rho_t\ret
& &-\alpha\int_Md\mu\,
(\partial_{i,1}f)g^{ij}_{\ 1}g^{k\ell}_{\ 2}
(\partial_{k,2}W)(\partial_{j,1}\partial_{\ell,2}W)\rho_t\ret
&=&\alpha\int_Md\mu\,(\partial_{i,1}f)g^{i\ell}_{\ 1}W
\frac{1}{\sqrt{{\rm det}\,g_2}}\partial_{j,2}
\sqrt{{\rm det}\,g_2}g^{jk}_{\ 2}
(\partial_{\ell,1}\partial_{k,2}W)\rho_t,
\label{I12}
\end{eqnarray}
where we have used the divergence theorem (\ref{GreenTh}). 
Recalling $W={\bf S}_1\cdot{\bf S}_2$, we have  
\begin{equation}
\partial_{\ell,1}\partial_{k,2}W
=\left(\partial_{\ell,1}{\bf S}_1\right)\cdot\left(\partial_{k,2}{\bf S}_2\right).
\end{equation}
Substituting this into the above result, we get 
\begin{eqnarray}
I_{12}&=&\alpha\int_Md\mu\,(\partial_{i,1}f)g^{i\ell}_{\ 1}
(\partial_{\ell,1}{\bf S}_1)W\cdot
\frac{1}{\sqrt{{\rm det}\,g_2}}\partial_{j,2}
\sqrt{{\rm det}\,g_2}g^{jk}_{\ 2}
\left(\partial_{k,2}{\bf S}_2\right)\rho_t\ret
&=&\alpha\int_Md\mu\,(\partial_{i,1}f)\mbox{\boldmath $\eta$}_1^iW\cdot
{\rm div}_2\left(\mbox{\boldmath $\eta$}_2\rho_t\right)\ret
&=&\alpha\int_Md\mu\,\frac{1}{\sqrt{{\rm det}\,g_1}}
\partial_{i,1}\sqrt{{\rm det}\,g_1}\mbox{\boldmath $\eta$}_1^ifW\cdot
{\rm div}_2\left(\mbox{\boldmath $\eta$}_2\rho_t\right)\ret
& &-\alpha\int_Md\mu\,f\frac{1}{\sqrt{{\rm det}\,g_1}}
\partial_{i,1}\sqrt{{\rm det}\,g_1}\mbox{\boldmath $\eta$}_1^iW\cdot
{\rm div}_2\left(\mbox{\boldmath $\eta$}_2\rho_t\right)\ret
&=&-\alpha\int_Md\mu\,f\,{\rm div}_1\left[\mbox{\boldmath $\eta$}_1W\cdot
{\rm div}_2\left(\mbox{\boldmath $\eta$}_2\rho_t\right)\right],
\label{I12res}
\end{eqnarray}
where $\mbox{\boldmath $\eta$}_\ell^i$ is given by (\ref{eta}).  
{From} (\ref{conseDeltat}), (\ref{conseF0}), (\ref{con11}), (\ref{con11f}), 
(\ref{I12}) and (\ref{I12res}), we obtain the Fokker-Planck equation,
\begin{eqnarray}
\frac{\partial \rho_t}{\partial t}&=&-\sum_\ell
{\rm div}_\ell\left(F_{0,\ell}\rho_t\right)
+\alpha\sum_\ell\left\{\Delta_\ell\rho_t-{\rm div}_\ell
\left[\xi_\ell\,{\rm div}_\ell(\xi_\ell\rho_t)\right]\right\}\ret
& &-\alpha\left\{{\rm div}_1\left[\mbox{\boldmath $\eta$}_1W
\cdot{\rm div}_2(\mbox{\boldmath $\eta$}_2\rho_t)\right]+
{\rm div}_2\left[\mbox{\boldmath $\eta$}_2W
\cdot{\rm div}_1(\mbox{\boldmath $\eta$}_1\rho_t)\right]\right\},
\label{FPeq0}
\end{eqnarray}
for $\alpha'=0$. 

Next consider the case with $\alpha'\ne 0$. To begin with, we note that 
\begin{eqnarray}
\av\left[\sigma_+^{(i)}\sigma_+^{(j)}\right]
&=&\frac{1}{2}\av\left[\left(\sigma_2^{(i)}+\sigma_1^{(i)}\right)
\left(\sigma_2^{(j)}+\sigma_1^{(j)}\right)\right]\ret
&=&\frac{1}{2}\left\{\av[\sigma_2^{(i)}\sigma_2^{(j)}]
+\av[\sigma_1^{(i)}\sigma_1^{(j)}]+\av[\sigma_2^{(i)}\sigma_1^{(j)}]
+\av[\sigma_1^{(i)}\sigma_2^{(j)}]\right\}\ret
&=&\frac{\alpha+\alpha'}{\Delta t}\delta^{ij}.
\end{eqnarray}
Similarly, 
\begin{equation}
\av\left[\sigma_-^{(i)}\sigma_-^{(j)}\right]
=\frac{\alpha-\alpha'}{\Delta t}\delta^{ij}.
\end{equation}
Further, we have 
\begin{eqnarray}
\av\left[\sigma_+^{(i)}\sigma_-^{(j)}\right]&=&
\frac{1}{2}\av\left[\left(\sigma_2^{(i)}+\sigma_1^{(i)}\right)
\left(\sigma_2^{(j)}-\sigma_1^{(j)}\right)\right]\ret
&=&\frac{1}{2}\left\{\av[\sigma_2^{(i)}\sigma_2^{(j)}]
-\av[\sigma_1^{(i)}\sigma_1^{(j)}]-\av[\sigma_2^{(i)}\sigma_1^{(j)}]
+\av[\sigma_1^{(i)}\sigma_2^{(j)}]\right\}\ret
&=&0.
\end{eqnarray}
Since we can write 
\begin{equation}
\av\left[\sigma_-^{(i)}\sigma_-^{(j)}\right]
=\frac{\alpha+\alpha'}{\Delta t}\delta^{ij}-\frac{2\alpha'}{\Delta t}\delta^{ij},
\end{equation}
it is sufficient to calculate the corrections from the second term 
in this right-hand side, with replacing $\alpha$ with $\alpha+\alpha'$ in 
the above result (\ref{FPeq0}). 

In (\ref{avFRFR}), the correction to 
$\av\left[g^{i\ell}_{\ 1}(\partial_{\ell,1}V_{\rm R})
g^{jk}_{\ 1}(\partial_{k,1}V_{\rm R})\right]$ is given by 
\begin{equation}
-\frac{4\alpha'}{\Delta t}\hat{\mbox{\boldmath $\zeta$}}_1^i
\cdot\hat{\mbox{\boldmath $\zeta$}}_1^j, 
\end{equation}
where $\hat{\mbox{\boldmath $\zeta$}}_\ell^i$ is given by (\ref{zeta}). 
Similarly, the correction to 
$\av\left[(\partial_{k,1}g^{ij}_{\ 1}\partial_{j,1}V_{\rm R})
(g^{k\ell}_{\ 1}\partial_{\ell,1}V_{\rm R})\right]$ in (\ref{avParFRFR}) 
is given by 
\begin{equation}
-\frac{4\alpha'}{\Delta t}\left(\partial_{k,1}\hat{\mbox{\boldmath $\zeta$}}_1^i\right)
\cdot\hat{\mbox{\boldmath $\zeta$}}_1^k. 
\end{equation}
Therefore the same calculations as those from (\ref{con11}) to (\ref{con11f}) 
yield the correction,  
\begin{equation}
-2\alpha'{\rm div}_1\left[\hat{\mbox{\boldmath $\zeta$}}_1
\cdot{\rm div}_1(\hat{\mbox{\boldmath $\zeta$}}_1\rho_t)\right],
\end{equation}
in the right-hand side of the Fokker-Planck equation (\ref{FPeq0}). 

In (\ref{avFR1FR2}), the correction to 
$\av\left[g^{i\ell}_{\ 1}(\partial_{\ell,1}V_{\rm R})
g^{jk}_{\ 2}(\partial_{k,2}V_{\rm R})\right]$ is given by 
\begin{equation}
-\frac{4\alpha'}{\Delta t}\hat{\mbox{\boldmath $\zeta$}}_1^i
\cdot\hat{\mbox{\boldmath $\zeta$}}_2^j. 
\end{equation}
Further, the correction to 
$\av\left[(\partial_{k,2}g^{ij}_{\ 1}\partial_{j,1}V_{\rm R})
(g^{k\ell}_{\ 2}\partial_{\ell,2}V_{\rm R})\right]$ in (\ref{avParFR1FR2}) 
is given by 
\begin{equation}
-\frac{4\alpha'}{\Delta t}\left(\partial_{k,2}\hat{\mbox{\boldmath $\zeta$}}_1^i\right)
\cdot\hat{\mbox{\boldmath $\zeta$}}_2^k. 
\end{equation}
Therefore similar calculations to those from (\ref{I12}) to (\ref{I12res}) 
yield the correction,  
\begin{equation}
-2\alpha'{\rm div}_1\left[\hat{\mbox{\boldmath $\zeta$}}_1
\cdot{\rm div}_2(\hat{\mbox{\boldmath $\zeta$}}_2\rho_t)\right],
\end{equation}
in the right-hand side of the Fokker-Planck equation (\ref{FPeq0}). 
In consequence, the Fokker-Planck equation is given by 
\begin{eqnarray}
\frac{\partial \rho_t}{\partial t}&=&-\sum_\ell
{\rm div}_\ell\left(F_{0,\ell}\rho_t\right)
+(\alpha+\alpha')\sum_\ell\left\{\Delta_\ell\rho_t-{\rm div}_\ell
\left[\xi_\ell\,{\rm div}_\ell(\xi_\ell\rho_t)\right]\right\}\ret
& &-(\alpha+\alpha')\left\{{\rm div}_1\left[\mbox{\boldmath $\eta$}_1W
\cdot{\rm div}_2(\mbox{\boldmath $\eta$}_2\rho_t)\right]+
{\rm div}_2\left[\mbox{\boldmath $\eta$}_2W
\cdot{\rm div}_1(\mbox{\boldmath $\eta$}_1\rho_t)\right]\right\}\ret
& &-2\alpha'\sum_{m,n}{\rm div}_m\left[\hat{\mbox{\boldmath $\zeta$}}_m
\cdot{\rm div}_n(\hat{\mbox{\boldmath $\zeta$}}_n\rho_t)\right].
\label{FPeq}
\end{eqnarray}

\Section{Derivation of the expansion (\ref{Jx1expand})}
\label{appendix:Jx1expand}

The metric $g_{ij,\ell}$ of $\sp^3$ is computed as 
\begin{equation}
g_{ij,\ell}=\left(\matrix{
1+\gamma_\ell{x_\ell^2}& \gamma_\ell{x_\ell y_\ell}& 
\gamma_\ell{x_\ell z_\ell}\cr
\gamma_\ell{y_\ell x_\ell}& 1+\gamma_\ell{y_\ell^2}&
\gamma_\ell{y_\ell z_\ell}\cr
\gamma_\ell{z_\ell x_\ell}& \gamma_\ell{z_\ell y_\ell}&
1+\gamma_\ell{z_\ell^2}\cr
}\right)
=\left(\matrix{
1+x_\ell^2& x_\ell y_\ell& x_\ell z_\ell\cr
y_\ell x_\ell& 1+y_\ell^2& y_\ell z_\ell\cr
z_\ell x_\ell& z_\ell y_\ell& 1+z_\ell^2\cr
}\right)+\cdots,
\end{equation}
where we have written 
\begin{equation}
\gamma_\ell=\frac{1}{\sqrt{1-r_\ell^2}}
\quad\mbox{with}\ \ r_\ell=\sqrt{x_\ell^2+y_\ell^2+z_\ell^2}.
\end{equation}
Therefore, the inverse $g^{ij}_{\ \ell}$ is given by 
\begin{equation}
g^{ij}_{\ \ell}=\left(\matrix{
1-x_\ell^2& -x_\ell y_\ell& -x_\ell z_\ell\cr
-y_\ell x_\ell& 1-y_\ell^2& -y_\ell z_\ell\cr
-z_\ell x_\ell& -z_\ell y_\ell& 1-z_\ell^2\cr
}\right)+\cdots.
\end{equation}
Using this, we have 
\begin{eqnarray}
(\partial_{x,1}W){\rm div}_1(\xi_1\rho)&=&
\frac{\partial {\bf S}_1\cdot{\bf S}_2}{\partial x_1}
\frac{1}{\sqrt{{\rm det}\,g_1}}\partial_{i,1}
\sqrt{{\rm det}\,g_1}g^{ij}_{\ 1}(\partial_{j,1}{\bf S}_1\cdot{\bf S}_2)\rho\ret
&=&-xg^{ij}_{\ 1}(\partial_{j,1}{\bf S}_1\cdot{\bf S}_2)\partial_{i,1}\rho+\cdots\ret
&=&x\left(x\frac{\partial \rho}{\partial x_1}+y\frac{\partial\rho}{\partial y_1}
+z\frac{\partial\rho}{\partial z_1}\right)+\cdots.
\label{DWdivxi1rho}
\end{eqnarray}
Similarly,
\begin{eqnarray}
W(\partial_{x,1}{\bf S}_1)\cdot{\rm div}_2(\mbox{\boldmath $\eta$}_2\rho)&=&
W(\partial_{x,1}{\bf S}_1)\cdot 
g^{ij}_{\ 2}(\partial_{j,2}{\bf S}_2)\partial_{i,2}\rho+\cdots\ret
&=&W(\partial_{x,1}\partial_{j,2}{\bf S}_1\cdot{\bf S}_2)g^{ij}_{\ 2}
\partial_{i,2}\rho+\cdots\ret
&=&W\left\{\partial_{j,2}\left[-x-\frac{1}{2}({\bf r}\cdot{\bf R})x_1+\cdots\right]\right\}
g^{ij}_{\ 2}\partial_{i,2}\rho+\cdots\ret
&=&Wg^{i1}_{\ 2}\partial_{i,2}\rho+
W\left[x_1x_2\frac{\partial\rho}{\partial x_2}+x_1y_2\frac{\partial\rho}{\partial y_2}
+x_1z_2\frac{\partial\rho}{\partial z_2}\right]+\cdots\ret
&=&\left(1-\frac{1}{2}r^2\right)\left[g^{11}_{\ 2}\frac{\partial\rho}{\partial x_2}
+g^{21}_{\ 2}\frac{\partial\rho}{\partial y_2}+g^{31}_{\ 2}
\frac{\partial\rho}{\partial z_2}\right]\ret
& &+\left[x_1x_2\frac{\partial\rho}{\partial x_2}+x_1y_2\frac{\partial\rho}{\partial y_2}
+x_1z_2\frac{\partial\rho}{\partial z_2}\right]+\cdots\ret
&=&\frac{\partial\rho}{\partial x_2}-\frac{1}{2}r^2\frac{\partial\rho}{\partial x_2}
-\left[x_2^2\frac{\partial\rho}{\partial x_2}+x_2y_2\frac{\partial\rho}{\partial y_2}
+x_2z_2\frac{\partial\rho}{\partial z_2}\right]\ret
& &+\left[x_1x_2\frac{\partial\rho}{\partial x_2}+x_1y_2\frac{\partial\rho}{\partial y_2}
+x_1z_2\frac{\partial\rho}{\partial z_2}\right]+\cdots\ret
&=&\frac{\partial\rho}{\partial x_2}-\frac{1}{2}r^2\frac{\partial\rho}{\partial x_2}
+x\left[x_2\frac{\partial\rho}{\partial x_2}+y_2\frac{\partial\rho}{\partial y_2}
+z_2\frac{\partial\rho}{\partial z_2}\right]+\cdots.
\label{WDS1diveta2rho}
\end{eqnarray}
We write 
\begin{equation}
\hat{\mbox{\boldmath $\zeta$}}_{i,\ell}
=\left(\zeta_{i,\ell}^{(1)},\zeta_{i,\ell}^{(2)},\zeta_{i,\ell}^{(3)}\right).
\end{equation}
Note that
\begin{eqnarray}
\zeta_{x,1}^{(a)}&=&\frac{\partial}{\partial x_1}
\left(S_1^{(0)}S_2^{(a)}-S_2^{(0)}S_1^{(a)}\right)\ret
&=&\frac{-x_1}{\sqrt{1-r_1^2}}S_2^{(a)}-\sqrt{1-r_2^2}
\frac{\partial S_1^{(a)}}{\partial x_1}.
\end{eqnarray}
Therefore, we have  
\begin{eqnarray}
\hat{\mbox{\boldmath $\zeta$}}_{x,1}&=&
\left(\frac{-x_1x_2}{\sqrt{1-r_1^2}}-\sqrt{1-r_2^2},\frac{-x_1y_2}{\sqrt{1-r_1^2}},
\frac{-x_1z_2}{\sqrt{1-r_1^2}}\right)\ret
&=&\left({-x_1x_2}-\sqrt{1-r_2^2},{-x_1y_2},{-x_1z_2}\right)+\cdots.
\label{zetax1}
\end{eqnarray}
In the same way, 
\begin{equation}
\hat{\mbox{\boldmath $\zeta$}}_{y,1}
=\left(-y_1x_2,-y_1y_2-\sqrt{1-r_2^2},-y_1z_2\right)+\cdots
\label{zetay1}
\end{equation}
and 
\begin{equation}
\hat{\mbox{\boldmath $\zeta$}}_{z,1}
=\left(-z_1x_2,-z_1y_2,-z_1z_2-\sqrt{1-r_2^2}\right)+\cdots.
\label{zetaz1}
\end{equation}
{From} these results, we obtain  
\begin{equation}
\hat{\mbox{\boldmath $\zeta$}}_{x,1}\cdot\hat{\mbox{\boldmath $\zeta$}}_{x,1}
=1-r_2^2+2x_1x_2+\cdots, 
\end{equation}
\begin{equation}
\hat{\mbox{\boldmath $\zeta$}}_{x,1}\cdot\hat{\mbox{\boldmath $\zeta$}}_{y,1}
=y_1x_2+x_1y_2+\cdots
\end{equation}
and 
\begin{equation}
\hat{\mbox{\boldmath $\zeta$}}_{x,1}\cdot\hat{\mbox{\boldmath $\zeta$}}_{z,1}
=z_1x_2+x_1z_2+\cdots. 
\end{equation}
Using these, we have 
\begin{eqnarray}
\hat{\mbox{\boldmath $\zeta$}}_{x,1}\cdot
{\rm div}_1(\hat{\mbox{\boldmath $\zeta$}}_1\rho)&=&
\hat{\mbox{\boldmath $\zeta$}}_{x,1}\cdot g^{ij}_{\ 1}
\hat{\mbox{\boldmath $\zeta$}}_{j,1}\partial_{i,1}\rho+\cdots\ret
&=&(1-r_2^2+2x_1x_2)g^{i1}_{\ 1}\partial_{i,1}\rho\ret
&+&(y_1x_2+x_1y_2)g^{i2}_{\ 1}\partial_{i,1}\rho
+(z_1x_2+x_1z_2)g^{i3}_{\ 1}\partial_{i,1}\rho+\cdots\ret
&=&(1-r_2^2+2x_1x_2)\left[(1-x_1^2)\frac{\partial\rho}{\partial x_1}
-x_1y_1\frac{\partial\rho}{\partial y_1}
-x_1z_1\frac{\partial\rho}{\partial z_1}\right]\ret
&+&(y_1x_2+x_1y_2)\frac{\partial\rho}{\partial y_1}
+(z_1x_2+x_1z_2)\frac{\partial\rho}{\partial z_1}+\cdots\ret
&=&\frac{\partial\rho}{\partial x_1}-r_2^2\frac{\partial\rho}{\partial x_1}
-x_1\left(x\frac{\partial\rho}{\partial x_1}+y\frac{\partial\rho}{\partial y_1}
+z\frac{\partial\rho}{\partial z_1}\right)\ret&+&
x_2\left(x_1\frac{\partial\rho}{\partial x_1}+y_1\frac{\partial\rho}{\partial y_1}
+z_1\frac{\partial\rho}{\partial z_1}\right)+\cdots.
\label{zeta1divzeta1rho}
\end{eqnarray}

In the same way, 
\begin{equation}
\hat{\mbox{\boldmath $\zeta$}}_{x,2}=
\left(x_1x_2+\sqrt{1-r_1^2},x_2y_1,x_2z_1\right)+\cdots,
\end{equation}
\begin{equation}
\hat{\mbox{\boldmath $\zeta$}}_{y,2}
=\left(y_2x_1,y_1y_2+\sqrt{1-r_1^2},y_2z_1\right)+\cdots
\end{equation}
and 
\begin{equation}
\hat{\mbox{\boldmath $\zeta$}}_{z,2}
=\left(z_2x_1,z_2y_1,z_1z_2+\sqrt{1-r_1^2}\right)+\cdots.
\end{equation}
Combining these, (\ref{zetax1}), (\ref{zetay1}) and (\ref{zetaz1}), we obtain   
\begin{equation}
\hat{\mbox{\boldmath $\zeta$}}_{x,1}\cdot\hat{\mbox{\boldmath $\zeta$}}_{x,2}
=-\left(1-\frac{1}{2}r_1^2-\frac{1}{2}r_2^2+2x_1x_2\right)+\cdots, 
\end{equation}
\begin{equation}
\hat{\mbox{\boldmath $\zeta$}}_{x,1}\cdot\hat{\mbox{\boldmath $\zeta$}}_{y,2}
=-2x_1y_2+\cdots
\end{equation}
and 
\begin{equation}
\hat{\mbox{\boldmath $\zeta$}}_{x,1}\cdot\hat{\mbox{\boldmath $\zeta$}}_{z,2}
=-2x_1z_2+\cdots. 
\end{equation}
Using these, we have 
\begin{eqnarray}
\hat{\mbox{\boldmath $\zeta$}}_{x,1}\cdot
{\rm div}_2(\hat{\mbox{\boldmath $\zeta$}}_2\rho)&=&
\hat{\mbox{\boldmath $\zeta$}}_{x,1}\cdot g^{ij}_{\ 2}
\hat{\mbox{\boldmath $\zeta$}}_{j,2}\partial_{i,2}\rho+\cdots\ret
&=&-\left(1-\frac{1}{2}r_1^2-\frac{1}{2}r_2^2+2x_1x_2\right)g^{i1}_{\ 2}
\partial_{i,2}\rho\ret
& &-2x_1y_2g^{i2}_{\ 2}\partial_{i,2}\rho-2x_1z_2g^{i3}_{\ 2}
\partial_{i,2}\rho+\cdots\ret
&=&-g^{i1}_{\ 2}\partial_{i,2}\rho
+\frac{1}{2}(r_1^2+r_2^2)\frac{\partial\rho}{\partial x_2}\ret
& &-2x_1\left(x_2\frac{\partial\rho}{\partial x_2}+y_2\frac{\partial\rho}{\partial y_2}
+z_2\frac{\partial\rho}{\partial z_2}\right)+\cdots\ret
&=&-\frac{\partial\rho}{\partial x_2}
+\frac{1}{2}(r_1^2+r_2^2)\frac{\partial\rho}{\partial x_2}\ret
&-&\left(\frac{3}{2}x+\frac{1}{2}X\right)
\left(x_2\frac{\partial\rho}{\partial x_2}+y_2\frac{\partial\rho}{\partial y_2}
+z_2\frac{\partial\rho}{\partial z_2}\right)+\cdots.
\label{zeta1divzeta2rho}
\end{eqnarray}
Substituting (\ref{DV0rho}), (\ref{DWdivxi1rho}), (\ref{WDS1diveta2rho}), 
(\ref{zeta1divzeta1rho}) and (\ref{zeta1divzeta2rho}) 
into (\ref{current}), 
we obtain the expansion (\ref{Jx1expand}).



\end{document}